\documentclass{emulateapj}
\shortauthors{Wang et al.}

\defcitealias{ta05}{TA05}

\begin{document}

\title{SDSS J102347.6$+$003841: A Millisecond Radio Pulsar Binary That Had
A Hot Disk During 2000-2001}

\author{Zhongxiang Wang\altaffilmark{1}, 
Anne M. Archibald\altaffilmark{1}, 
John R. Thorstensen\altaffilmark{2}, 
Victoria M. Kaspi\altaffilmark{1}, 
Duncan R. Lorimer\altaffilmark{3},
Ingrid Stairs\altaffilmark{4,5},
Scott M. Ransom\altaffilmark{6}
}

\altaffiltext{1}{\footnotesize Department of Physics,
Ernest Rutherford Physics Building,
McGill University, 3600 University Street, Montreal, QC H3A 2T8, Canada}

\altaffiltext{2}{Department of Physics and Astronomy, Dartmouth College,
6127 Wilder Laboratory, Hanover, NH 03755, USA}

\altaffiltext{3}{Physics Department, West Virginia University, 
210 Hodges Hall, Morgantown, WV 26506, USA}

\altaffiltext{4}{Department of Physics and Astronomy, 
University of British Columbia,
6224 Agricultural Road, Vancouver, BC V6T 1Z1, Canada}

\altaffiltext{5}{Centre for Astrophysics \& Supercomputing,
Swinburne University of Technology,
Mail 39 PO Box 218, Hawthorn Vic 3122, Australia}

\altaffiltext{6}{National Radio Astronomy Observatory,
520 Edgemont Rd., Charlottesville, VA 22901, USA}

\begin{abstract}
The Sloan Digital Sky Survey (SDSS) source J102347.6$+$003841 
was recently revealed to be a binary 1.69 millisecond radio pulsar 
with a 4.75 hr orbital period and a $\sim$0.2 $M_{\sun}$ companion.  
Here we analyze the SDSS spectrum 
of the source in detail.  The spectrum was taken on 2001 February 1, when 
the source was in a bright state and showed broad, double-peaked hydrogen 
and helium lines --- dramatically different from the G-type absorption 
spectrum seen from 2002 May onward.  The lines are consistent with emission 
from a disk around the compact primary.  We derive properties of 
the disk by fitting the SDSS continuum with a simple disk model, and 
find a temperature range of 2000--34000~K from the outer to inner edge 
of the disk.  The disk inner and outer radii were approximately 10$^9$ and
5.7$\times 10^{10}$ cm, respectively.  These results further emphasize
the unique feature of the source: it is a system likely at 
the end of its transition from an X-ray binary to a recycled radio pulsar. 
The disk mass is estimated to have been $\sim 10^{23}$ g, most of which 
would have been lost due to pulsar wind ablation (or due to 
the propeller effect if the disk had extended inside the light cylinder of
the pulsar) before the final disk disruption event. 
The system could undergo repeated episodes of disk formation. 
Close monitoring of the source is needed to catch the system in 
its bright state again, so that this unusual
example of a pulsar-disk interaction can be studied in much finer detail.

\end{abstract}


\section{INTRODUCTION}

The Sloan Digital Sky Survey (SDSS) source J102347.6$+$003841 
(hereafter J1023) is located well off the Galactic plane 
at $b=45.8^{\circ}$.  At $V = 17.5$ and $K_s = 15.9$, it is bright enough
to have been detected by several optical and near-infrared sky surveys 
since 1950 (e.g., \citealt{zac+04, 2mass}). \citet{bond+02} first drew 
attention to this source following its detection in the Faint Images of 
the Radio Sky at Twenty Centimeters (FIRST) survey \citep{bwh95}.
Their follow-up optical observations revealed flickering and a spectrum showing 
strong, double-peaked hydrogen and helium emission lines.  Based
on these, they suggested that the system was a cataclysmic variable 
(CV), that is, a binary system in which a white-dwarf primary accretes
from a close companion through Roche-lobe overflow. 
Around the same time, optical spectra obtained by SDSS and 
\citet{szk+03} appeared similar to those found by \citet{bond+02}.

It was therefore surprising when Thorstensen \& Armstrong (2005; 
hereafter \citetalias{ta05})
obtained further spectra, starting in 2003 January, that showed 
substantially lower flux and an ordinary-looking G-type spectrum,
with {\it no} detectable emission lines (see also \citealt{hom+06}, who
found the same G-type spectrum in 2002 May).
\citet{wou+04} published time-series photometry
that showed a smooth, $\sim 0.3$ mag modulation at a 4.75-hour period; 
\citetalias{ta05} confirmed this result, and also found 
that the radial velocity of the G-type spectrum varies on the
same period with an amplitude of  $268 \pm 4$ km s$^{-1}$.
It has thus been established that J1023 is a binary and 
the G-type spectrum is from the companion star in the binary.
Furthermore, on the basis of the radial velocity and results from 
light-curve fitting, \citetalias{ta05} inferred that the compact primary in 
J1023 was likely to be a neutron star or possibly a black hole, and  
suggested that the system resembles a Low Mass X-ray Binary (LMXB) 
rather than a CV.  X-ray emission from the system was subsequently detected by
\citet{hom+06}, but the X-ray luminosity is low, 
$L_{\rm X} \approx 10^{32}$ ergs s$^{-1}$ for source distance $D= 1.3$ kpc.  
This is considerably short of the $\sim 2$  L$_{\odot}$ compact object 
luminosity needed to account for the heating effect \citepalias{ta05}.

The puzzles set forth by TA05 were resolved when  
Archibald et al. (2009) discovered pulsed radio emission from J1023
in an untargeted radio pulsar survey carried out using 
the Green Bank Telescope in mid-2007, thereby confirming 
that J1023 is a neutron star binary. Moreover, the pulsar's spin period 
$P$ is only 1.69 ms, making it the fifth fastest among $\sim$170 known 
millisecond pulsars (MSPs; $P\lesssim 30$ ms; e.g., \citealt{man+05,lor08}). 

The discovery of the millisecond pulsar makes the bright spectrum 
observed during 2000--2001 extremely interesting. The double-peaked
emission lines are commonly seen in CVs and LMXBs, and are a 
typical feature of accretion disks.  This suggests that during the period 
in which emission-line spectra were observed, the companion star 
overflowed its Roche lobe and mass transfer occurred, forming a disk 
surrounding the MSP. It is not clear whether significant mass accreted 
onto the neutron star, because no evidence of strong, 
accretion-powered X-ray emission has been found. In any case,
by showing both an accretion disk and radio pulsations, this system 
is likely the first such binary found at the end of its transition from
an LMXB to a radio MSP.
It is a valuable laboratory for studying the evolution of 
X-ray binaries and the formation of radio MSPs.

In this paper, we report the results of the first detailed analysis of 
the SDSS spectrum.  We summarize
in \S~\ref{sec:targ} the 
binary properties of J1023, estimated from previous optical 
and recent radio observations (\citetalias{ta05}; Archibald et al. 2009). 
In \S~\ref{sec:obs} we briefly describe
the SDSS spectroscopic observation of J1023,
and present our study of spectral features
in \S~\ref{sec:spec}. From this, we derive the properties of
the short-lived disk. We discuss the implications of the results
in \S~\ref{sec:disc}.
\begin{center}
%%\plotone{f1.eps}
\includegraphics[scale=0.7]{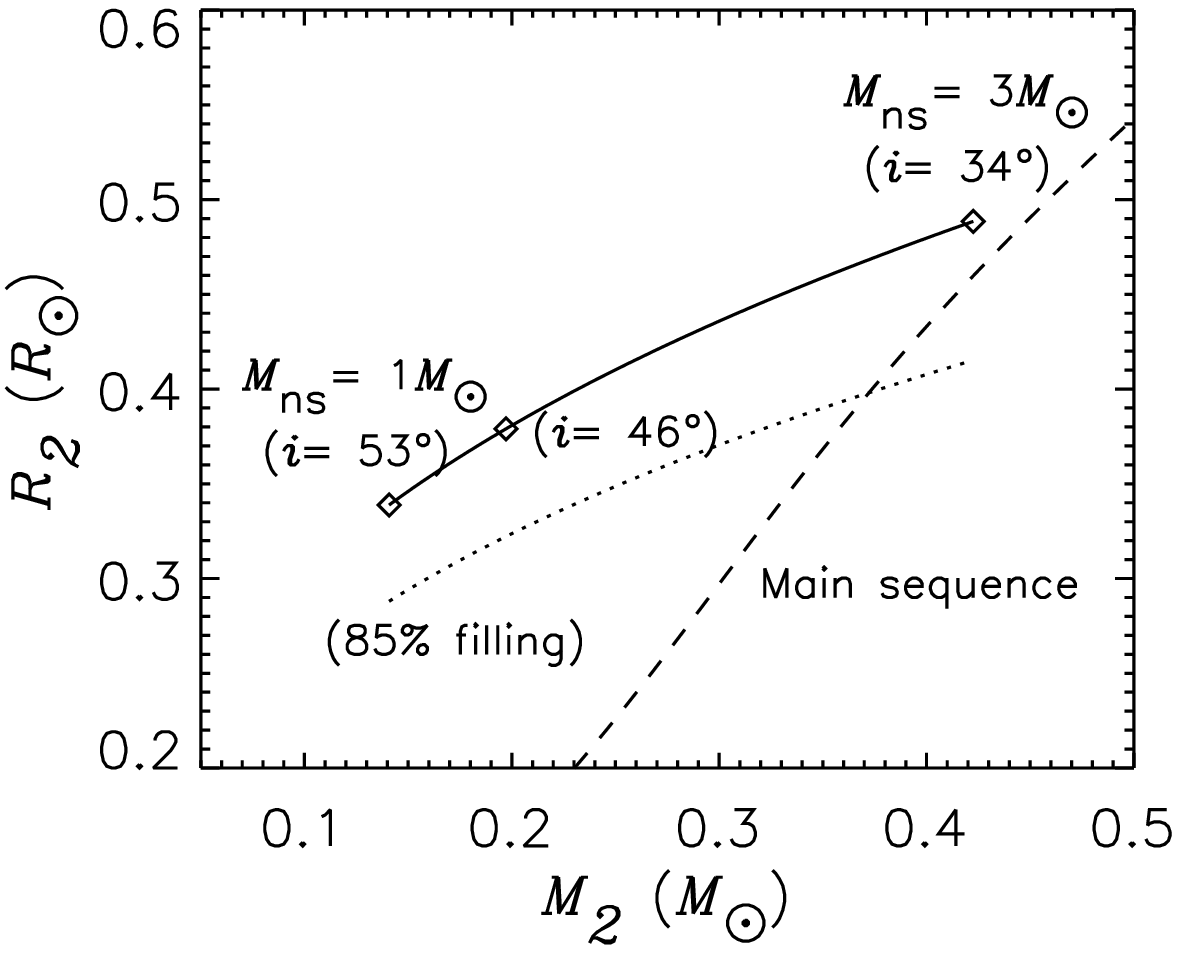}
\figcaption{Mass and radius values (solid curve) derived for
the companion star in J1023 when $M_{\rm ns}= 1$--3 $M_{\sun}$ is considered. 
Several $i$ values are marked on the curve (diamonds). 
For $M_{\rm ns}=1.4\ M_{\sun}$, $M_2=0.2 M_{\sun}$ and $i=46\arcdeg$.
The mass-radius relation for low-mass main-sequence 
stars \citep{tout96} is plotted (dashed curve), indicating that
main-sequence stars cannot fill in the Roche lobe. 
For illustration, we also show the radius values for the companion star when
the radius is only 85\% that of the Roche lobe (dotted curve).
\label{fig:mr} }
\end{center}

\subsection{Binary Properties}
\label{sec:targ}
From radio timing of the pulsar in J1023, the pulsar's Keplerian mass function
is found to be 1.1$\times 10^{-3}\ M_{\sun}$.  Combining this with 
the measured radial velocity amplitude of the companion star, 
the mass ratio $q$ is determined,  $q=M_{\rm ns}/M_2=7.1\pm0.1$. Here
$M_{\rm ns}$ and $M_2$ are, respectively, the neutron star and 
companion masses.  For a canonical neutron star mass 
$M_{\rm ns}=1.4\ M_{\sun}$, this implies $M_2\simeq 0.2\ M_{\sun}$.
If we allow a wider range of $M_{\rm ns}$ from 1 to 3 $M_{\sun}$, then
$M_2 = 0.14$--0.42 $M_{\sun}$.
Accordingly, the orbital inclination angle $i$ is limited to 
34$\arcdeg$--$53\arcdeg$, nearly identical to the range of inclinations
estimated from the heating effect by \citetalias{ta05}.
If in addition the radius of the companion star $R_2$ 
is equal to that of its Roche lobe\footnote{From studies of MSP binaries, 
the companion stars have often been found to be close to filling 
their Roche lobes (\citetalias{ta05}; \citealt{sta01,rey+07}).}, 
then the companion star's radius is almost entirely a function of its 
mass.  Figure~\ref{fig:mr} shows the derived mass and radius values for the
companion star.
The \citetalias{ta05} light curve fits showed the companion to have 
an effective temperature $T_{\rm eff}$ of  
5600--5700~K, similar to that of a mid G-type dwarf.  
Based on the discussion above, we adopt canonical values of  
$M_{\rm ns}=1.4\ M_{\sun}$ and $M_2=0.2\ M_{\sun}$
throughout this paper, unless mentioned otherwise.

\section{SDSS SPECTROSCOPY}    % Section 2 
\begin{figure*}
\begin{center}
%%\plotone{f2.eps}
\includegraphics[scale=0.65]{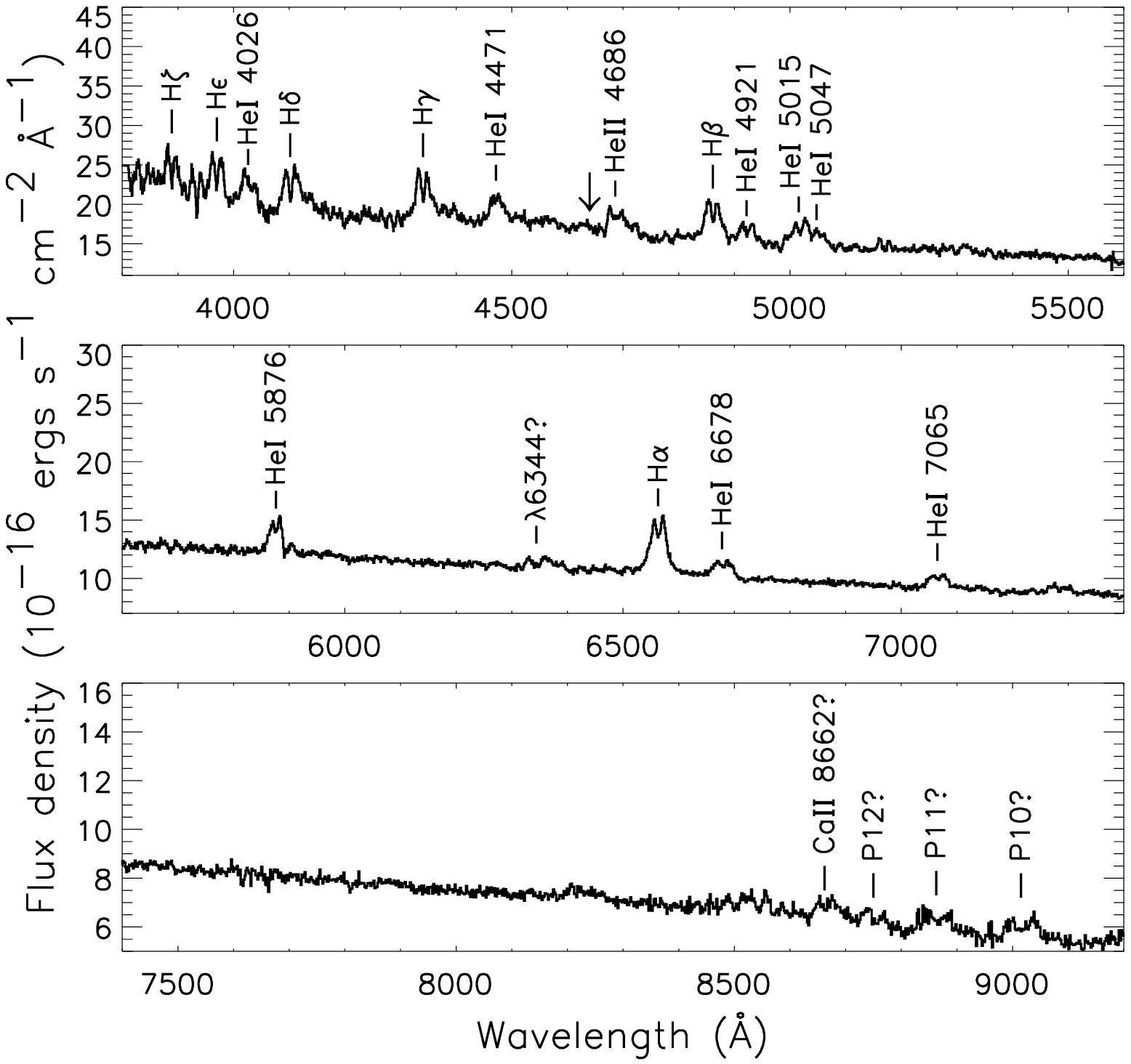}
\figcaption{SDSS spectrum of J1023. The emission lines in the spectrum
are labelled. The arrow in the top panel indicates the absence of 
the $\lambda$~4640 \AA\ emission feature. This feature is commonly seen
in LMXBs (e.g., \citealt{vm95}), in which it arises from X-ray irradiated 
companion stars.
\label{fig:spec} }
\end{center}
\end{figure*}

\label{sec:obs}
The spectrum of J1023 was taken 
on 2001 February 1 using
the dedicated SDSS 2.5-m telescope at Apache Point Observatory
\citep{york+00}, and was included in the 7th SDSS data 
release \citep{dr7+08}. 
The spectrum covers from 3800 to 9200 \AA\ with 
spectral resolution $\lambda/\Delta\lambda\sim$2000;  the total
on-source time was over 1 hr.
It was reduced using the SDSS Spectro2d pipeline
(version v5\_3\_12; details can be found on the SDSS webpages,
e.g., http://www.sdss.org/dr7/products/spectra/index.html).
In the pipeline flux calibration procedure, 
a few standard stars are
observed on each spectroscopic plate, and
flux calibration is achieved by comparing the spectra of the standard stars
to their model spectra.  Generally, the 3$\sigma$ uncertainty of 
SDSS spectra in $r'$ filter is 0.15 mag. The SDSS spectrum of J1023
is shown in Figure~\ref{fig:spec}. The wavelengths are vacuum heliocentric.

\section{SPECTRAL ANALYSIS} 
\label{sec:spec}

\subsection{Emission Lines}

The SDSS spectrum shows emission lines, 
mostly from the hydrogen Balmer series and He I, on a 
smooth, blue continuum.  All the emission lines are double-peaked.
The spectrum resembles those of compact LMXBs and CVs; in those systems,
the optical emission 
arises mostly in an accretion disk, and the double-peaked profiles 
reflect the line-of-sight components of the disk rotation
velocity (e.g., \citealt{hm86}).
The Bowen fluorescence blend near $\lambda$4640, which is
commonly seen in LMXBs, is notably absent.
This emission feature consists of \ion{N}{3}, \ion{C}{3}, and \ion{O}{2}
lines (e.g., \citealt{km80, sfk89}) and is thought to arise from
X-ray irradiated companion stars (e.g., \citealt{deguchi85, sc02}).
The absence of the $\lambda$4640 feature suggests
the lack of X-ray irradiated, highly ionized gas in the system and
probably indicates a low X-ray luminosity at the time. This is consistent
with expectations given the X-ray flux upper limit of the source 
[$F_{\rm 2-10\ keV}\lesssim 1.2\times 10^{-10}$ ergs s$^{-1}$ cm$^{-2}$, 
derived from the \textit{Rossi X-ray Timing Explorer (RXTE)} All-Sky 
Monitor data; Archibald et al. 2009].
LMXBs that show the Bowen emission feature typically have 
large X-ray luminosities, $\sim$10$^{36}$ erg s$^{-1}$.

For each emission line, we
measured the total flux and the wavelengths of the red and blue peaks in
the following manner.
We first dereddened the spectrum using the 
wavelength-dependent extinction law from \citet{fit99}  and taking 
$A_V=0.14$ from \citet*{sfd98}. 
We then estimated the continuum with a 6-order polynomial.
To fit the continuum-subtracted double-peaked lines, we used either
two-Gaussian or two-Lorentz functions.  For most of the lines,  
the two-Lorentz profile gave the better fit.  Tables~\ref{tab:bal} 
and \ref{tab:he} give
the results for the hydrogen and helium lines, respectively. 
The vacuum wavelength values used are from
the atomic spectra database at the National Institute of Standards and 
Technology (NIST).\footnote{see http://physics.nist.gov/PhysRefData/ASD/lines\_form.html} 
Because of the crowding of the lines 
in the blue region, it is difficult to determine the continuum flux exactly;
in measuring those lines, we estimated the continuum from the polynomial fit.

\begin{figure*}
%%\plotone{f3.eps}
\begin{center}
\includegraphics[scale=0.6]{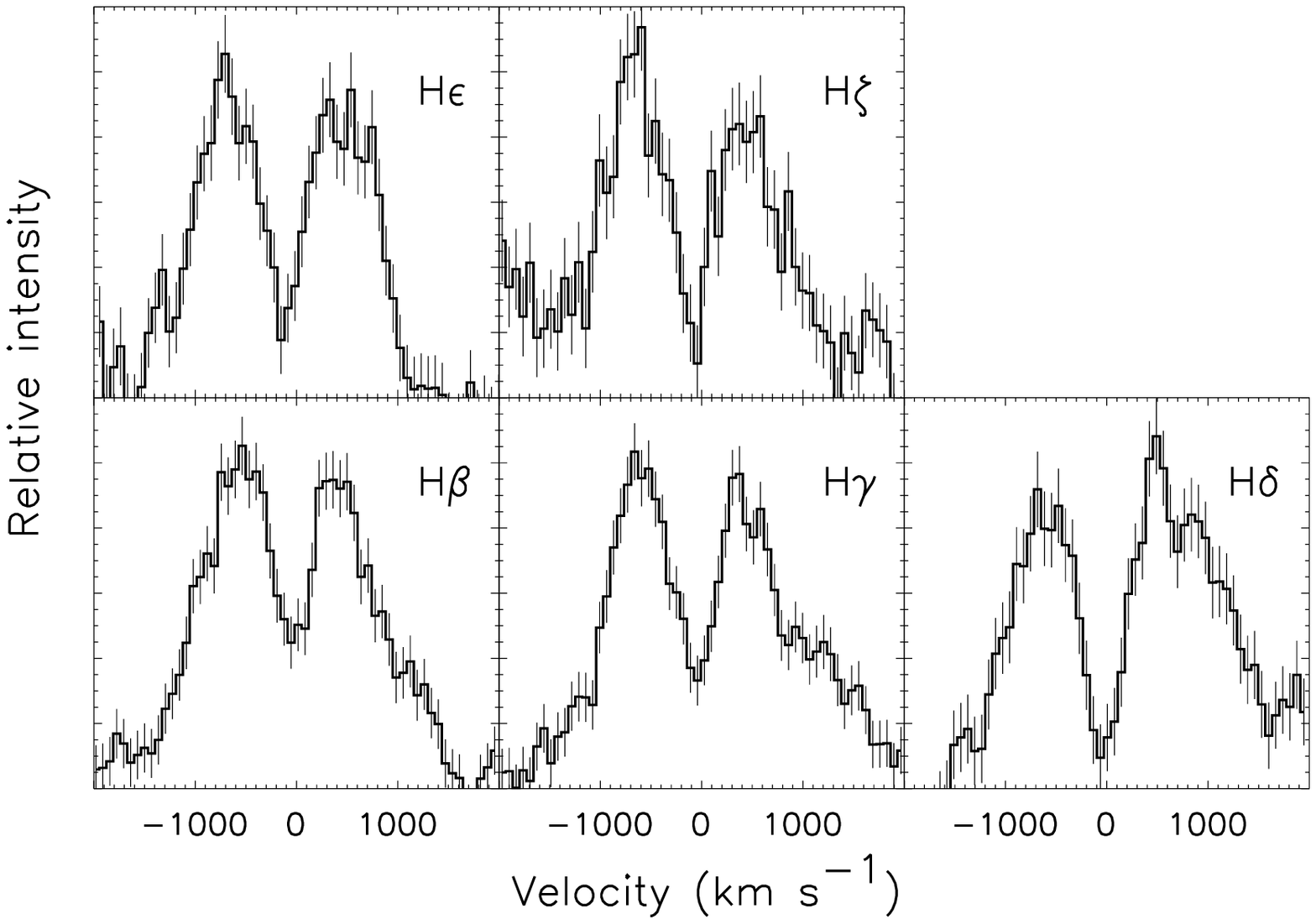}
\figcaption{Balmer H$\beta$--H$\zeta$ emission lines in the SDSS spectrum.
The blue- and red-shifted peaks are not symmetric.
The right wing of the red-shifted H$\delta$ peak may 
be contaminated by another line component.
\label{fig:hline} }
\end{center}
\end{figure*}

We focus here on the hydrogen lines, since they are stronger than the 
helium lines and thus more accurately measurable.  As can be seen, the 
blue and red peaks of the H$\beta$,
$\gamma$, $\epsilon$, and $\zeta$ lines have velocities in
the ranges of $-$(610--710) and (470--490) km s$^{-1}$
(Figure~\ref{fig:hline}).  
The average values are $-660$ km s$^{-1}$ and 
480 km s$^{-1}$. Comparing to these four lines, H$\delta$ has 
a stronger, more red-shifted profile, while the blue and red peak velocities 
of H$\alpha$ are significantly smaller. The blue and red peaks of
the lines are not symmetric, suggesting contributions from other 
components.  A hot spot on a disk, arising from the interaction 
between the gas flow from a companion star and the disk, 
can cause such asymmetry (e.g., \citealt{oro+94}; \citealt{mas+00}). 

The fluxes of the Balmer lines are comparable with each other, 
indicating that the disk was optically thick \citep{wil80}.
For example, the peak intensity (in frequency units) ratio of 
H$\alpha$ to H$\beta$ is $\simeq$1.7. This value is lower than that 
found in the classical case of an optically thin gas 
(e.g., \citealt{ost74}). In addition, the central-valley
regions of the Balmer lines clearly have a shape of `V', not `U' 
(Figures~\ref{fig:hline} and \ref{fig:ha}), also 
indicating an optically thick disk \citep{hm86}. The double-peaked shape 
of the lines is generally considered to be a function of several 
disk parameters \citep{smak81,hm86}. 
The line intensity, $F(u)$, can be written as
\begin{equation}
F(u) \propto \int_{r_1}^{r_z}\frac{r^{3/2-\alpha}dr}{(1-u^2r)^{1/2}}
	[1+4\Upsilon^2(i)u^2r(1-u^2r)]^{1/2},
\end{equation}
where $r$ is the normalized disk radius ($r=1$ corresponds to the
outer edge of the disk), $u$ is the dimensionless radial velocity ($u=1$ at
$r=1$), $r_1$ is the ratio of the inner disk radius $r_{\rm in}$ to 
the outer disk radius $r_{\rm out}$, $r_z$=min(1, $u^{-2}$), and 
$\Upsilon = \sin i\tan i$ \citep{smak81,hm86,oro+94}.
In the equation, a power-law model, $f(r)\propto r^{-\alpha}$, is assumed
for the disk density function of the emitting atoms, where $\alpha$ is usually
found to have values between 1.0--2.0.

We used this profile function to fit the Balmer lines in the SDSS spectrum.
The free parameters are peak velocity $V_p$, $i$, $r_1$, and $\alpha$.  
Because the lines are obviously not identical and 
there are relatively large flux uncertainties in the blue region, 
we fit the H$\alpha$ line mainly to constrain the parameters.
Although the two H$\alpha$ peaks are not symmetric, the 
minimum $\chi^2$ value is 
113 for 108 degrees of freedom, indicating that the disk profile 
fits the H$\alpha$ line reasonably well.
The fitting is sensitive to $\alpha$: $\alpha \simeq 1.6$ 
always provides the best fit. $V_p$ is tightly constrained in the narrow range 
of 395($-$20, +10) km s$^{-1}$ (99\% confidence). 
Because changes of $r_1$ only cause 
small disk area and flux changes and $i$ is limited by the central valley 
region, only a small fraction of the whole line profile, 
$r_1$ and $i$ cannot be tightly constrained. 
The 99\% confidence ranges are $r_1 \leq 0.04$ and $i\leq 44\arcdeg$
(note that since the inner disk radius cannot be smaller than that of 
the neutron star, $r_1$ has a lower limit, $r_1\gtrsim 10^{-5}$). 
We tested fitting H$\beta$ and found different results:
$V_p=540\pm30$ km s$^{-1}$, $r_1=0.0002$--0.08, and $i\leq 60\arcdeg$ 
(99\% confidence; $\alpha\simeq$1.6).
The deep valley in the H$\beta$ line and other bluer hydrogen lines 
favors a large inclination angle \citep{hm86}. However, 
inclination angles determined from fitting double-peaked lines obviously are  
not reliable. For example, the hydrogen lines in the SDSS spectrum 
have different 
central depths, which would suggest different inclination angles.
Also, in other cases, the central valley of a line from the same source
was seen to be variable 
(e.g., \citealt{oro+94}).
\begin{center}
%%\plotone{f4.eps}
\includegraphics[scale=0.6]{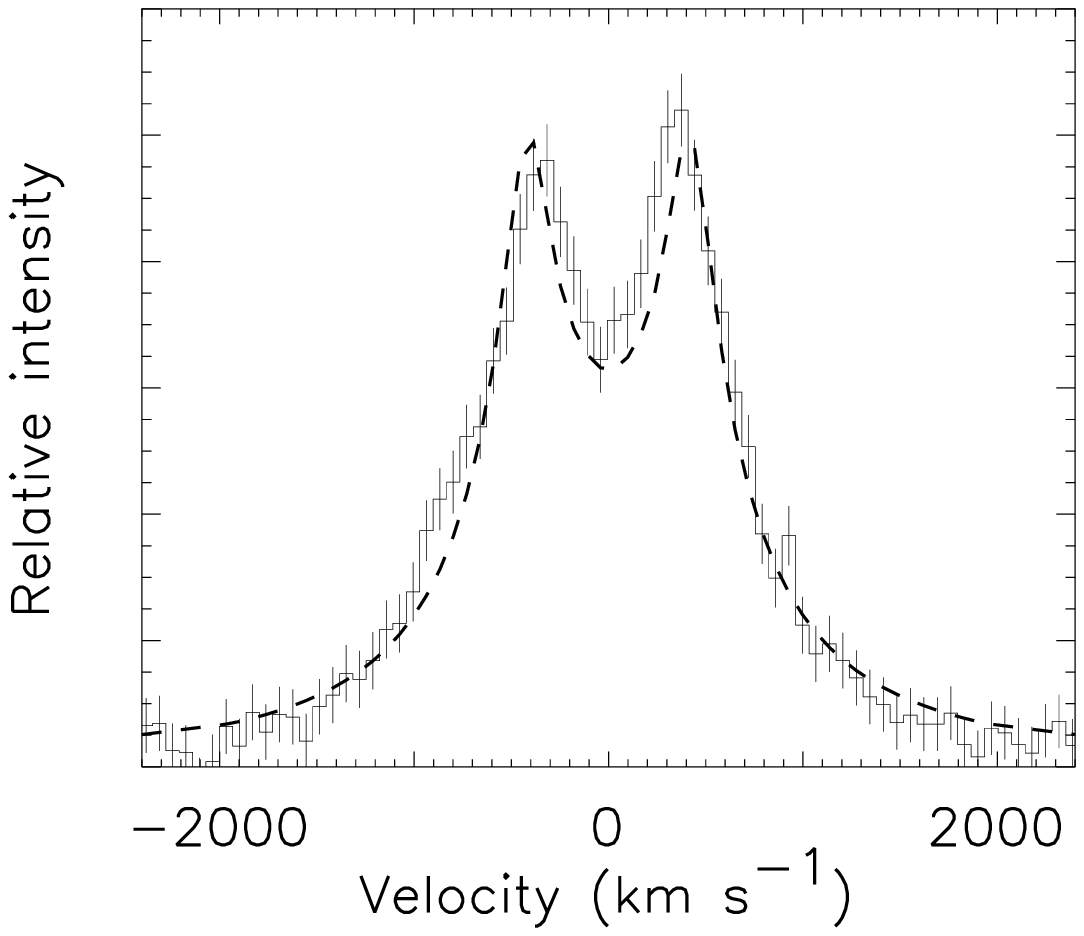}
\figcaption{H$\alpha$ emission line in the SDSS spectrum. 
The disk model profile with $\alpha=1.6$, $V_p=395$ km s$^{-1}$,
$r_1=0.004$, and $i=40\arcdeg$, is plotted as the dashed curve.
Because of the asymmetry in the two peaks, 
the model profile does not fit the peak regions well. 
\label{fig:ha} }
\end{center}

If we assume the disk is Keplerian, 
the disk outer edge rotation 
velocity $V_d$ and $r_{\rm out}$ can be inferred to be 
$V_d=V_p/\sin i$ and 
$r_{\rm out} = (GM_{\rm ns}/V_d^2)\simeq 1.1$--1.3$\times 10^{11} \sin^2(i)$ cm.
The Roche lobe radius $R_1$ of the neutron star can be found and the disk 
is expected to be cut off at the tidal radius 
$R_{\rm tides}=0.9 R_1 = 5.7\times 10^{10}$ cm (for $M_2=0.2\ M_{\sun}$ and
$M_{\rm ns}=1.4\ M_{\sun}$).
Since $r_{\rm out}$ cannot be larger than $R_{\rm tides}$,
we find $i \leq 41\arcdeg$--46$\arcdeg$, consistent with the low inclination 
angle value set by the mass ratio. On the other hand, this implies that 
$r_{\rm out}\simeq R_{\rm tides}$, because when $M_{\rm ns}=1.4\ M_{\sun}$ is
assumed,
$i$ should not be significantly different from 46$\arcdeg$.
We checked the case when $M_{\rm ns}=2.9\ M_{\sun}$. The resulting $i$
is between 31$\arcdeg$--$34\arcdeg$, also approximately consistent with
the value given above in \S~\ref{sec:targ}.
The larger peak velocities of
other hydrogen lines probably suggest that they were emitted from 
a smaller disk area.
For example, for $V_p=540$ km s$^{-1}$ (H$\beta$ peak velocity)
and $i=46\arcdeg$, 
$r_{\rm out}\approx 3.3\times 10^{10}$ cm.

The peak velocities of the hydrogen lines have negative mean values,
which would suggest that the disk (and the neutron star) was moving 
towards us. The velocities of the lines are significantly different.
If we consider only H$\gamma$, $\epsilon$, and $\zeta$ lines, which
have similar velocity values, the average radial velocity is 
$-72\pm36$ km s$^{-1}$, not well determined. 
This velocity would be orbital because \citetalias{ta05}
found nearly zero radial velocity for the binary system.
Considering the radio timing measurements
of the system and the orbital phase 0.65--0.92 (phase 0.0 corresponds to
the ascending node of the pulsar) for the SDSS observation,
the radial velocity of the neutron star was in the range of 
($-$22)--($+$33) km s$^{-1}$ during the observation. The median velocity 
value is $\sim$10 km s$^{-1}$. These possible radial velocity values are
different from the observed value, suggesting that either the lines
were distorted by emission from other components such as a hot spot, or 
the disk was not axis-symmetric (e.g., \citealt{mas+00} and references therein).
Generally, radial velocity curves derived
from emission lines from a disk may not be a reliable indicator for
the primary's orbital motion
(e.g., \citealt{oro+94}; \citealt{mas+00}). 

The helium lines in the SDSS spectrum are relatively strong.
For example, the flux ratio between H$\alpha$ and \ion{He}{1} $\lambda$6680
is approximately 3.3, lower than typical values found in normal
CV systems (e.g., \citealt{tho+02}).  It is likely that helium is enhanced
\citep{wf82}.  As can be seen in Figure~\ref{fig:mr}, the companion must 
be {\it much} less massive than a main-sequence star of the same radius.
Also, it is {\it much} hotter than a main-sequence star in 
the allowed mass range. While J1023 is not a CV, it is similar in that 
it is a semi-detached binary; as a general rule, secondary stars in CVs tend 
to be slightly {\it cooler} than they `should' be based on main-sequence 
expectations \citep{kniggedonor}.
The secondary in J1023 is therefore highly anomalous.  
A few CVs do show similar `too-hot' secondary stars 
\citep{thoreipsc, thorqzser}; in those cases, the secondary
has evidently undergone significant nuclear evolution before 
losing large amounts of mass, leaving an undermassive core enriched in helium.
The secondary in J1023
appears to be a similarly evolved, stripped core, much hotter, and much larger
in radius, than one would expect given its mass.
The star probably had a mass of $\gtrsim 1 M_{\sun}$ at onset of mass 
transfer \citep{del08}, and has lost much of its mass during 
its X-ray binary phase. 

\subsection{Continuum}
\label{sec:con}

The SDSS spectrum is substantially brighter than those spectra obtained 
after 2002 May; to show this, its continuum 
is plotted with the spectrum from \citetalias{ta05} in Figure~\ref{fig:disk}.
The latter spectrum, also dereddened with $A_V=0.14$,  
is a mean of 23 spectra obtained in 2003--2004.
Based on the spectra, a mid-G spectral type has been identified
for the low-mass companion star in J1023.
The average $V$ magnitude of the companion is 17.5 \citepalias{ta05},
corresponding to a dereddened flux of 
4.3$\times 10^{-16}$ ergs$^{-1}$ s$^{-1}$ cm$^{-2}$ \AA$^{-1}$. 
In Figure~\ref{fig:disk}, we plot a G5V spectrum smoothed and 
normalized to the flux value at $V$ wavelength;
the spectrum is from the stellar 
spectral flux library published by \citet{pic98}. As can be seen,
the library spectrum is generally consistent with that of the quiescent emission
from J1023---the blue-end 
region of the former is lower than that of the latter, but the flux 
uncertainties of the latter in the region are as large as 15\%.  
A more stringent test of how well the spectrum matches a G5 dwarf could 
be made if we had a flux-calibrated
spectrum of J1023 covering wavelengths to 3500 \AA , as a sharp
turnover is expected at 4500 \AA.
In the red end, if the G5V spectrum were extended to near-infrared $JHK_s$ 
wavelengths, it would have flux values approximately consistent 
with the 2MASS measurements of J1023 
($J=16.3\pm 0.1$, $H=15.7\pm0.1$, $K_s=15.9$;
the uncertainty on $K_s$ could not be determined, and the observations
were carried out on 2000 February 6).  All in all, the
evidence supports the use of the G5V spectrum
for the companion star. 

Including the normalized G5V spectrum, we fit the SDSS continuum with a simple 
disk model. The flux at wavelength $\lambda$ from the disk is
\begin{equation}
F_{\lambda} = \frac{4\pi hc^2\cos i}{\lambda^5 D^2}
\int_{r_{\rm in}}^{r_{\rm out}} \frac{rdr}{e^{hc/\lambda kT_d} - 1}\ ,
\end{equation}
where $h$ is the Planck constant, $c$ is the speed of light, and $k$ is
the Boltzmann constant.
We assumed that the disk temperature $T_d$ is 
a power-law function of the disk radius $r$, $T_d\propto r^{-p}$, 
and further considered $p= 3/4$, the same as that in the steady thin disk
model \citep{fkr92}. 
If the neutron star had strong 
X-ray emission at the time due to accretion, which is unlikely based on 
the X-ray upper limit (Archibald et al. 2009) and absence of the Bowen 
fluorescence feature, 
$p= 3/7$ might be assumed \citep{vrt+90}.
Under the simple model, the more likely case would be that $p$ varied 
during the disk lifetime.
If that is the case, multiple observations of a future disk phase 
would be needed to detect spectral changes caused by disk evolution.

We first tried three parameters in the fitting: $D$, $r_{\rm in}$,
and the temperature $T_d^0$ at $r_{\rm in}$, while
$r_{\rm out}$ was set to be 
$r_{\rm out}=0.9 R_1\approx 5.7\times 10^{10}$ cm 
(for $M_{\rm ns}=1.4\ M_{\sun}$ and $i = 46\arcdeg$).
The distance $D$ is constrained by the flux from the companion,
since it is a function of $(R_2/D)^2$. For example, if we assume 
a G5V dwarf ($T_{\rm eff}\simeq 5600$ K, 
radius $R_{\ast}=0.92\ R_{\sun}$, and absolute magnitude $M_V=5.1$) 
as the companion, $V=17.5$ would imply a distance of 2.8 kpc.
When $R_2$ is assumed to be equal to the Roche-lobe radius 
($0.38\ R_{\sun}$) of the companion, $D\simeq 1.2$ kpc, inferred from 
$(R_2/D)=(0.92 R_{\sun}/2.8$ kpc).
Similarly, \citetalias{ta05} have found 
$D\simeq (2.2\ {\rm kpc})(M_2/M_{\sun})^{1/3}$, and $D\simeq 1.3$ kpc when
$M_2=0.2 M_{\sun}$.
Therefore, we searched for the best fit in the range of $D= 0.9$--1.7 kpc.
\begin{center}
%%\plotone{f5.eps}
\includegraphics[scale=0.6]{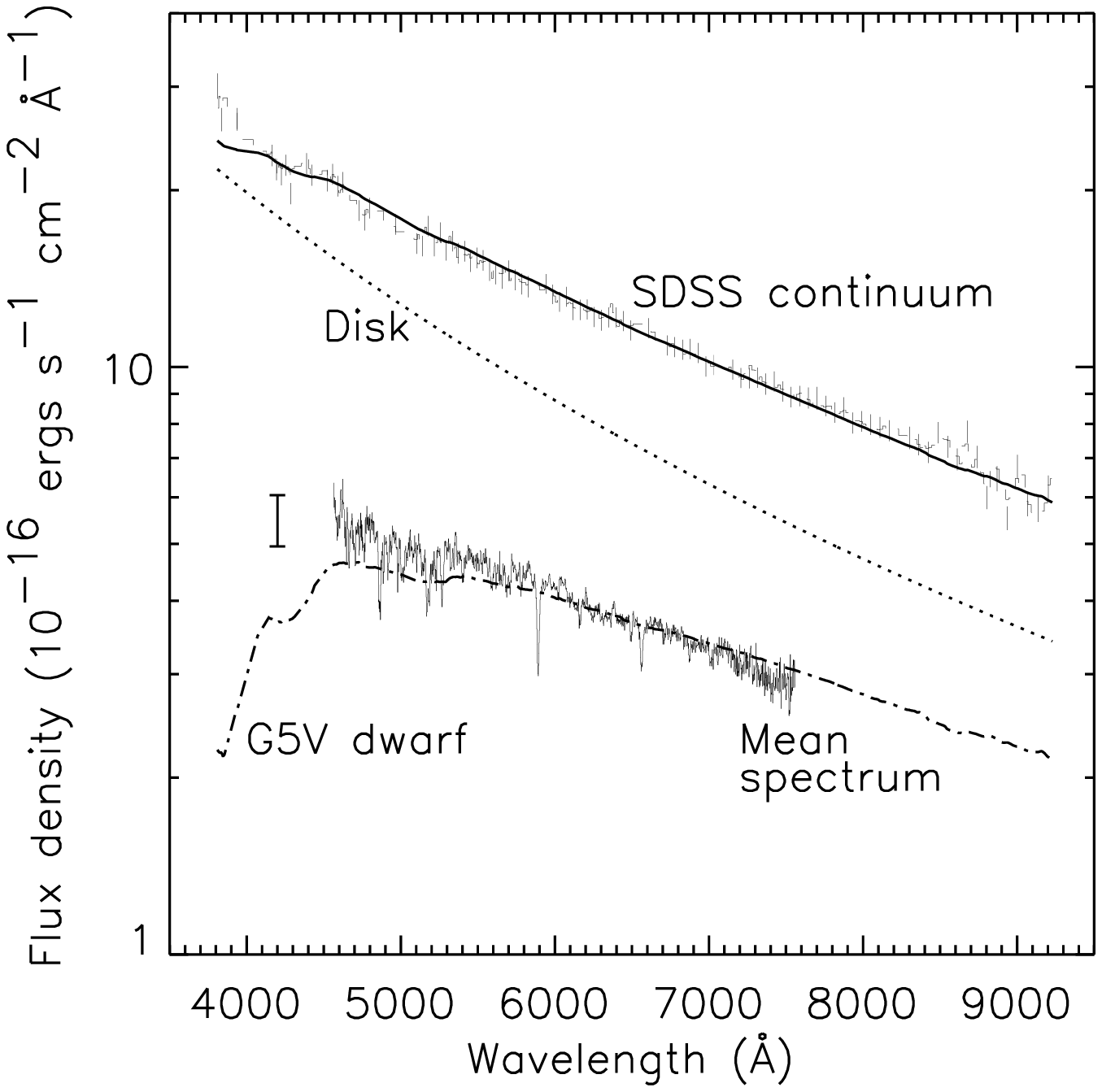}
\figcaption{
Comparison of the SDSS continuum from 2001 
(top light dashed histogram over-plotted with flux uncertainties short lines) 
with the mean J1023 spectrum (bottom solid curve) 
obtained in 2003--2004.  The latter has substantially lower flux,
and can be described by a G5V spectrum (dash-dotted curve). The error bar
at the left side indicates the large uncertainties in
the blue-end region of the mean spectrum. Combining
the G5V spectrum with that of a disk model (dotted curve), the summed spectrum
(top dark solid curve) fits the SDSS spectrum well, supporting the existence
of a disk during 2000--2001.
\label{fig:disk} }
\end{center}

Optical emission from J1023 is modulated because the visible area of 
the heated face of the companion star varies as a function of orbital
phase \citepalias{ta05}. We assumed this
flux modulation the same in the bright state as in the quiescent,
which is expected 
since no strong X-ray emission (to heat the companion) was seen. 
The SDSS observation
was made at orbital phase 0.65--0.92, right within the flat top of 
the companion's modulated light curve (\citetalias{ta05}; note that because 
phase 0.0 is defined differently here from in \citetalias{ta05}, the phase
in this paper leads by $1/4$). The average $V$ magnitude during the phase range 
is 17.354$\pm0.014$, implying a 15\% flux increase at the time from 
the companion. 
We simply increased the flux of the G5V spectrum by 15\%. 
Because the flux increases are
caused by the heated half surface of the companion star,
they are actually wavelength-dependent (TA05). However,
since the SDSS spectrum is much brighter than that of the companion,  
the differences are negligible in the fitting.

The flux uncertainties on the SDSS spectrum are approximately 2\% in the middle
region and 4\% at the end regions. If we use the uncertainties, 
the minimum 
reduced $\chi^2$ (2922 degrees of freedom) resulting from the fitting is 4.6. 
We therefore included a systematic uncertainty, which on average could be 
$\sim$5\% in the $r'$ band. In addition, it should also contain uncertainties
on the companion's spectrum, since no uncertainties for the
G5V library spectrum were assumed.
This systematic uncertainty was added in quadrature,
while its value was adjusted to have the minimum reduced $\chi^2=1$. 
A value of 3.4\% for the uncertainty was required.
The fitting is sensitive to the parameters, and we found 
$D\simeq 1.0$ kpc, $r_{\rm in}\simeq 1.5\times 10^9$ cm,
and $T_d^0\simeq 34000$ K. 
The $r_{\rm in}$ value is consistent with those 
from our H$\alpha$ fitting, while $D$ is $\sim$20\% lower than those derived
above and by \citetalias{ta05}. The $D$ value may suggest that $R_2$
was not equal to, but lower than the Roche lobe radius.
If $D= 1.3$ kpc is required, the spectrum is less well fit, and
$r_{\rm in}\simeq 1.4\times 10^{9}$ cm and $T_d^0\simeq 40000$ K.
In Figure~\ref{fig:disk}, the best-fit model spectrum is shown.

We further considered that this was a steady thin disk case.
Given the values obtained above, the mass accretion rate $\dot{M}$ in the disk 
can be estimated from $T_d \simeq 8000 \dot{M}_{16}^{1/4} r_{10}^{-3/4}$ K
\citep{fkr92}, $\dot{M}_{16}\simeq 1.1$, where $r_{10}$ and $\dot{M}_{16}$ 
are $r$ and $\dot{M}$ in units of 10$^{10}$ cm and 10$^{16}$ g s$^{-1}$,
respectively.
Based on this $\dot{M}$ value, the Alfv\'{e}n radius $r_{\rm M}$,
which is expected to be the inner disk radius, is much smaller than 
the obtained $r_{\rm in}$.
$r_{\rm M}\sim 2.5\times 10^6 (\dot{M}_{16}/1.1)^{-2/7}
(B/10^{8}\ {\rm G})^{4/7}$ cm, where $B$ is the surface magnetic field strength
of the neutron star ($B\sim 10^8$ G; see \S~\ref{sec:disc} below).
The discrepancy indicates inconsistency in the results 
when the steady thin disk case is considered. 
We therefore tested to find the alternative best 
fit by setting $r_{\rm in}\sim R_{\Omega}\simeq 2.4\times 10^{6}$ cm 
(where $R_{\Omega}$ is the corotation radius) and 
$T_d\propto \dot{M}^{1/4}r^{-3/4}$.
Generally, when $r_{\rm in}\sim R_{\Omega}$, accretion onto a neutron star
occurs and the pulsed radio emission from the pulsar is quenched 
due to accretion.
It can be noted that to satisfy $r_{\rm in}\sim r_{\rm M}\sim R_{\Omega}$, 
$\dot{M}_{16}\sim 1$.  Since $r_{\rm M}$ is weakly dependent 
on $\dot{M}$, a large range of $\dot{M}$ values will result in small changes 
of $r_{\rm M}$ (or $r_{\rm in}$), indicating that it is reasonable to fix
$r_{\rm in}$ at the $R_{\Omega}$ value.
To search for a good fit, 
$r_{\rm out}$ was set to be a free parameter
in the reasonable range: $r_{\rm out}\leq 5.7\times 10^{10}$ cm.
We found from the fitting,
$\dot{M}\simeq 1.0\times 10^{16}$ g s$^{-1}$ (which is consistent with the
assumed $r_{\rm in}$ value), $r_{\rm out}= 5.7\times 10^{10}$ cm, 
and $D=1.1$ kpc, but the minimum reduced $\chi^2\simeq 1.9$.
In order to have $\chi^2\simeq 1$,
a systematic uncertainty of 5\% would have to be included in the fitting. 

\section{Discussion}
\label{sec:disc}

We analyzed the SDSS spectrum, which was obtained 
in 2001 February when J1023 was in a bright state.
Although the double-peaked hydrogen lines in the spectrum 
have shapes more complex than those in a standard disk profile, 
our detailed study of them shows that their properties are consistent 
with those of a disk around the neutron star in J1023.
Considering a G5V spectrum normalized to the brightness of the companion star,
we have also found that the SDSS continuum can be fit with a simple disk 
model, supporting its disk origin. 
These studies thus help indicate that in the bright state a disk existed 
in the binary, demonstrating the important feature
implied by the source: at the beginning of a radio MSP life, its companion star 
in the binary is still able to overflow its Roche lobe and a disk can 
thus be formed around the MSP.

From fitting the continuum, the temperature profile of the disk is
estimated to be $9100 r_{10}^{-3/4}$~K.
Such a temperature profile implies
$\dot{M}\simeq 1.1\times 10^{16}$ g s$^{-1}$ and a small,
$\sim 2.5\times 10^6$ cm Alfv\'{e}n radius when the standard steady 
thin disk is assumed. The radius value is much smaller than the
$r_{\rm in}$ value obtained from the fitting, indicating inconsistency 
in the results.  However, it can be shown that given the derived $\dot{M}$ 
value, the viscous (or radial drift) timescale could be 
as long as $\sim$110 days (the viscosity 
coefficient $\alpha =0.1$ is assumed; \citealt{fkr92}) for the outer disk. 
The starting time of the disk formation is constrained by the SDSS imaging 
observations, which were made on 1999 March 22 and indicate that J1023 was in
the quiescent state at the time 
(e.g., $g'=17.99$, $r'=17.43$; \citealt{bond+02,szk+03}).
The disk formed at some time later but before 2000 May 6 when 
the first bright-state spectrum was taken by \citet{bond+02}, 
and had existed for $>8$ month but $<2$ yr before the SDSS spectroscopic 
observation.
Because the viscous timescale is comparably long, these constraints 
suggest that the disk could have 
had $r_{\rm in}\sim 10^9$ cm if it had not fully extended
to the allowed smallest radius.
We note that $B$ currently is estimated to be low, $B\sim$10$^8$ G 
(although the spin-down rate $\dot{P}$ of the pulsar is quite uncertain,
$\dot{P}\sim 7\times 10^{-21}$ s s$^{-1}$; \citealt{arc+09}), supporting
a small Alfv\'{e}n radius.

Our assumption of a standard thin disk could be too 
simplified. On the other hand, it is plausible that the SDSS spectrum was 
not well calibrated 
in flux---the real spectrum would be steeper and 
thus $r_{\rm in}$ could be as small as
$\sim 2.5\times 10^6$ cm ($\dot{M}_{16}=1.0$ and $B= 10^8$ G are used).
This $r_{\rm in}$ value would be comparable to $R_{\Omega}$, suggesting that
accretion onto the pulsar could have occurred. However, the $\dot{M}$ value 
would imply an accretion luminosity of 
$L_{\rm X}\simeq GM_{\rm ns}\dot{M}/R_{\rm ns} \simeq 2\times 10^{36}$ ergs s$^{-1}$.
Comparing this luminosity to the X-ray flux upper limit, 
either the source is $>$10 kpc
away or no accretion onto the neutron star occurred. Since 
multiple lines of evidence point at a distance of $\sim$1.3 kpc,
the latter is the likely case. 

The light cylinder (LC) radius of the MSP in J1023 is 
$r_{\rm lc}\simeq 8.1\times 10^6$ cm. Generally, if 
$R_{\Omega} \ll r_{\rm in} < r_{\rm lc}$, the system would have been
in the propeller phase \citep{is75} and the radio pulsar would have 
been quenched by the accretion flow (e.g., \citealt{cam+98}). The minimum
luminosity $L_{\rm lc}$ produced from the accretion flow would have been 
when $r_{\rm in}\sim r_{\rm lc}$, 
$L_{\rm lc}\simeq GM_{\rm ns}\dot{M}/r_{\rm lc}
\simeq 2\times 10^{35} \dot{M}_{16}$ erg s$^{-1}$. 
For $D=1.3$ kpc, 
$F_{\rm min}\simeq 10^{-9} \dot{M}_{16}$\ erg s$^{-1}$ cm$^{-2}$. 
This flux would have 
been detected by the \textit{RXTE} All-Sky Monitor. 
The upper limit on $\dot{M}$ given by the X-ray flux
upper limit is 10$^{15}$ g s$^{-1}$. Therefore, if the $\dot{M}$ value we 
have derived is approximately correct, $r_{\rm in}$ would have been larger
than $r_{\rm lc}$ and the radio pulsar would not have been
quenched at the time. This is consistent with the large $r_{\rm in}$ value
we obtained. On the other hand, if the $\dot{M}$ value is overestimated, 
the disk could have extended inside the LC and quenched the radio pulsar.  

We estimate the total disk mass $M_d$ by simply using the standard, 
$\alpha$-prescription disk model and integrating the surface 
density $\Sigma$ over the disk area. 
In the model,
$\Sigma\simeq 5.7 \alpha^{-4/5} \dot{M}_{16}^{7/10} r_{10}^{-3/4}$ g cm$^{-2}
\simeq 36 (\alpha/0.1)^{-4/5} \dot{M}_{16}^{7/10} r^{-3/4}_{10}$ g cm$^{-2}$. 
Using the inner and outer disk radii obtained above,
$M_d\approx 1.7\times 10^{23}$ g. The formation time
would have been 0.5 yr if the mass transfer rate had been constant and equal to 
$10^{16}$ g s$^{-1}$. This timescale is consistent with
the constraints on the disk formation time set by the optical observations.

The existence of a disk for more than a year makes J1023 an interesting
case for studying the pulsar-disk interaction. According to the standard
pulsar accretion scenario (\citealt{cam+98} and references therein),
the minimum mass accretion rate $\dot{M}_{\rm min}$ in a disk, 
which is required for the disk to be stable against
a pulsar's radiation pressure, is obtained when $r_{\rm M}\sim r_{\rm lc}$.
Detailed calculations made by \citet{ea05} show that
a disk can still be stable if its inner radius is only slightly 
larger than $r_{\rm lc}$. 
For the J1023 MSP, 
$\dot{M}_{\rm min}\sim 2\times 10^{14}$ g s$^{-1}$. With such a low rate,
the disk in J1023 would have lost most ($\sim$95\%) of its mass before 
the final disk disruption. The mass loss rate of the disk would have been 
3$\times 10^{15}$ g s$^{-1}$ on average, where 2.0 yr is assumed for
the disk lifetime. 
Because the spin-down luminosity of the pulsar is probably as large as
$\sim 5\times 10^{34}$ erg s$^{-1}$ (from the light curve fitting,
\citetalias{ta05} have found that the required luminosity from the MSP 
is $\sim 10^{34}$ erg s$^{-1}$, supporting the spin-down luminosity value),
the mass loss could have been due to the pulsar wind ablation 
of the disk \citep{mh01}.
On the other hand, if the disk had extended inside the LC and quenched 
the pulsar wind, the mass loss would have been caused by the propeller effect: 
the mass inflow is halted by the magnetic field outside of the corotation
radius and the material is repelled away from the system.

It is likely that J1023 will repeat its process of having 
an outflow and forming a disk, although the timescale between two such events
is not known. If we do see the next event in the near future, it would 
provide a great 
opportunity for studying the pulsar-disk interaction. The short lifetime
and brightness of the disk would allow close multiwavelength monitoring,
from which we might be able to test current pulsar accretion theory.
For example, we might find at which stage of the disk evolution
pulsed radio emission is quenched.  Would it be, as generally suggested, 
when the disk extends right inside the LC?
If the transition to the propeller phase occurs, we might be able to 
seek direct evidence for mass loss due to the propeller 
effect (e.g., \citealt{rom+09}).  
On the other hand, we would study the evolution of an accretion disk 
as its being ablated by a pulsar wind. 
More importantly we might directly observe 
how a disk is disrupted by a pulsar.

\acknowledgements
    Funding for SDSS and SDSS-II has been provided by the Alfred P. Sloan 
Foundation, the Participating Institutions, the National Science Foundation, 
the U.S. Department of Energy, the National Aeronautics and Space 
Administration, the Japanese Monbukagakusho, the Max Planck Society, 
and the Higher Education Funding Council for England. 
The SDSS Web Site is http://www.sdss.org/.

  The SDSS is managed by the Astrophysical Research Consortium for 
the Participating Institutions. The Participating Institutions are 
the American Museum of Natural History, Astrophysical Institute Potsdam, 
University of Basel, University of Cambridge, Case Western Reserve University,
University of Chicago, Drexel University, Fermilab, the Institute for 
Advanced Study, the Japan Participation Group, Johns Hopkins University, 
the Joint Institute for Nuclear Astrophysics, the Kavli Institute for 
Particle Astrophysics and Cosmology, the Korean Scientist Group, 
the Chinese Academy of Sciences (LAMOST), Los Alamos National Laboratory, 
the Max-Planck-Institute for Astronomy (MPIA), the Max-Planck-Institute 
for Astrophysics (MPA), New Mexico State University, Ohio State University, 
University of Pittsburgh, University of Portsmouth, Princeton University, 
the United States Naval Observatory, and the University of Washington.

This research was supported by NSERC via Discovery Grants
and by the FQRNT and CIFAR. 
VMK holds a Canada Research Chair and
the Lorne Trottier Chair in Astrophysics \& Cosmology, and is a
Fellow of the Royal Society of Canada. 
JRT acknowledges support from the U.S. National Science Foundation,
through grants AST-0307413 and AST-0708810.
IHS acknowledges support from 
the Swinburne University of Technology Visiting Distinguished 
Researcher Scheme.

%%{\it Facility:} \facility{Gemini:South}

\bibliographystyle{apj}
%%\bibliography{msp}

\begin{thebibliography}{42}
\expandafter\ifx\csname natexlab\endcsname\relax\def\natexlab#1{#1}\fi

\bibitem[{{Abazajian} {et al.}(2009)}]{dr7+08}
{Abazajian}, K., et al. 2009, \apjs, 182, 543 

\bibitem[{Archibald} {et al.}(2009)]{arc+09}
Archibald, A. M., et al. 2009, Science, 324, 1411

\bibitem[{{Becker} {et~al.}(1995){Becker}, {White}, \& {Helfand}}]{bwh95}
{Becker}, R.~H., {White}, R.~L., \& {Helfand}, D.~J. 1995, \apj, 450, 559

\bibitem[{{Bond} {et~al.}(2002){Bond}, {White}, {Becker}, \&
  {O'Brien}}]{bond+02}
{Bond}, H.~E., {White}, R.~L., {Becker}, R.~H., \& {O'Brien}, M.~S. 2002,
  \pasp, 114, 1359

\bibitem[{{Campana} {et~al.}(1998){Campana}, {Colpi}, {Mereghetti}, {Stella},
  \& {Tavani}}]{cam+98}
{Campana}, S., {Colpi}, M., {Mereghetti}, S., {Stella}, L., \& {Tavani}, M.
  1998, \aapr, 8, 279

\bibitem[{{Deguchi}(1985)}]{deguchi85}
{Deguchi}, S. 1985, \apj, 291, 492

\bibitem[{{Deloye}(2008)}]{del08}
{Deloye}, C.~J. 2008, in American Institute of Physics Conference Series, Vol.
  983, 40 Years of Pulsars: Millisecond Pulsars, Magnetars and More, ed.
  C.~{Bassa}, Z.~{Wang}, A.~{Cumming}, \& V.~M. {Kaspi}, 501--509

\bibitem[{{Ek{\c s}{\.I}} \& {Alpar}(2005)}]{ea05}
{Ek{\c s}{\.I}}, K.~Y. \& {Alpar}, M.~A. 2005, \apj, 620, 390

\bibitem[{{Fitzpatrick}(1999)}]{fit99}
{Fitzpatrick}, E.~L. 1999, \pasp, 111, 63

\bibitem[{{Frank} {et~al.}(1992){Frank}, {King}, \& {Raine}}]{fkr92}
{Frank}, J., {King}, A., \& {Raine}, D. 1992, {Accretion Power in
  Astrophysics.} (Camb.~Astrophys.~Ser., Vol.~21,)

\bibitem[{{Homer} {et~al.}(2006){Homer}, {Szkody}, {Chen}, {Henden}, {Schmidt},
  {Anderson}, {Silvestri}, \& {Brinkmann}}]{hom+06}
{Homer}, L., {Szkody}, P., {Chen}, B., {Henden}, A., {Schmidt}, G., {Anderson},
  S.~F., {Silvestri}, N.~M., \& {Brinkmann}, J. 2006, \aj, 131, 562

\bibitem[{{Horne} \& {Marsh}(1986)}]{hm86}
{Horne}, K. \& {Marsh}, T.~R. 1986, \mnras, 218, 761

\bibitem[{{Illarionov} \& {Sunyaev}(1975)}]{is75}
{Illarionov}, A.~F. \& {Sunyaev}, R.~A. 1975, \aap, 39, 185

\bibitem[{{Kallman} \& {McCray}(1980)}]{km80}
{Kallman}, T. \& {McCray}, R. 1980, \apj, 242, 615

\bibitem[{{Knigge}(2006)}]{kniggedonor}
{Knigge}, C. 2006, \mnras, 373, 484

\bibitem[{Lorimer}(2008)]{lor08}
{Lorimer}, D. R. 2008, Living Reviews in Relativity, 11, 8

\bibitem[{{Manchester} {et~al.}(2005){Manchester}, {Hobbs}, {Teoh}, \&
  {Hobbs}}]{man+05}
{Manchester}, R.~N., {Hobbs}, G.~B., {Teoh}, A., \& {Hobbs}, M. 2005, \aj, 129,
  1993

\bibitem[{{Mason} {et~al.}(2000){Mason}, {Skidmore}, {Howell}, {Ciardi},
  {Littlefair}, \& {Dhillon}}]{mas+00}
{Mason}, E., {Skidmore}, W., {Howell}, S.~B., {Ciardi}, D.~R., {Littlefair},
  S., \& {Dhillon}, V.~S. 2000, \mnras, 318, 440

\bibitem[{{Miller} \& {Hamilton}(2001)}]{mh01}
{Miller}, M.~C. \& {Hamilton}, D.~P. 2001, \apj, 550, 863

\bibitem[{{Orosz} {et~al.}(1994){Orosz}, {Bailyn}, {Remillard}, {McClintock},
  \& {Foltz}}]{oro+94}
{Orosz}, J.~A., {Bailyn}, C.~D., {Remillard}, R.~A., {McClintock}, J.~E., \&
  {Foltz}, C.~B. 1994, \apj, 436, 848

\bibitem[{{Osterbrock}(1974)}]{ost74}
{Osterbrock}, D.~E. 1974, {Astrophysics of Gaseous Nebulae} (Research supported
  by the Research Corp., Wisconsin Alumni Research Foundation, John Simon
  Guggenheim Memorial Foundation, Institute for Advanced Studies, and National
  Science Foundation.~San Francisco, W.~H.~Freeman and Co., 1974)

\bibitem[{{Pickles}(1998)}]{pic98}
{Pickles}, A.~J. 1998, \pasp, 110, 863

\bibitem[{{Reynolds} {et~al.}(2007){Reynolds}, {Callanan}, {Fruchter},
  {Torres}, {Beer}, \& {Gibbons}}]{rey+07}
{Reynolds}, M.~T., {Callanan}, P.~J., {Fruchter}, A.~S., {Torres}, M.~A.~P.,
  {Beer}, M.~E., \& {Gibbons}, R.~A. 2007, \mnras, 379, 1117

\bibitem[{{Romanova} {et~al.}(2009){Romanova}, {Lovelace}, {Ustyugova}, \&
  {Koldoba}}]{rom+09}
{Romanova}, M.~M., {Lovelace}, R.~V.~E., {Ustyugova}, G.~V., \& {Koldoba},
  A.~V. 2009, ArXiv e-prints

\bibitem[{{Schachter} {et~al.}(1989){Schachter}, {Filippenko}, \&
  {Kahn}}]{sfk89}
{Schachter}, J., {Filippenko}, A.~V., \& {Kahn}, S.~M. 1989, \apj, 340, 1049

\bibitem[{{Schlegel} {et~al.}(1998){Schlegel}, {Finkbeiner}, \&
  {Davis}}]{sfd98}
{Schlegel}, D.~J., {Finkbeiner}, D.~P., \& {Davis}, M. 1998, \apj, 500, 525

\bibitem[{{Skrutskie} {et~al.}(2006)}]{2mass}
{Skrutskie}, M.~F. {et~al.} 2006, \aj, 131, 1163

\bibitem[{{Smak}(1981)}]{smak81}
{Smak}, J. 1981, Acta Astronomica, 31, 395

\bibitem[{{Stappers} {et~al.}(2001){Stappers}, {van Kerkwijk}, {Bell}, \&
  {Kulkarni}}]{sta01}
{Stappers}, B.~W., {van Kerkwijk}, M.~H., {Bell}, J.~F., \& {Kulkarni}, S.~R.
  2001, \apjl, 548, L183

\bibitem[{{Steeghs} \& {Casares}(2002)}]{sc02}
{Steeghs}, D. \& {Casares}, J. 2002, \apj, 568, 273

\bibitem[{{Szkody} {et~al.}(2003)}]{szk+03}
{Szkody}, P. {et~al.} 2003, \aj, 126, 1499

\bibitem[{{Thorstensen} \& {Armstrong}(2005)}]{ta05}
{Thorstensen}, J.~R. \& {Armstrong}, E. 2005, \aj, 130, 759

\bibitem[{{Thorstensen} {et~al.}(2002{\natexlab{a}}){Thorstensen}, {Fenton},
  {Patterson}, {Kemp}, {Halpern}, \& {Baraffe}}]{thorqzser}
{Thorstensen}, J.~R., {Fenton}, W.~H., {Patterson}, J., {Kemp}, J., {Halpern},
  J., \& {Baraffe}, I. 2002{\natexlab{a}}, \pasp, 114, 1117

\bibitem[{{Thorstensen} {et~al.}(2002{\natexlab{b}}){Thorstensen}, {Fenton},
  {Patterson}, {Kemp}, {Krajci}, \& {Baraffe}}]{tho+02}
{Thorstensen}, J.~R., {Fenton}, W.~H., {Patterson}, J.~O., {Kemp}, J.,
  {Krajci}, T., \& {Baraffe}, I. 2002{\natexlab{b}}, \apjl, 567, L49

\bibitem[{{Thorstensen} {et~al.}(2002{\natexlab{c}}){Thorstensen}, {Fenton},
  {Patterson}, {Kemp}, {Krajci}, \& {Baraffe}}]{thoreipsc}
---. 2002{\natexlab{c}}, \apjl, 567, L49

\bibitem[{{Tout} {et~al.}(1996){Tout}, {Pols}, {Eggleton}, \& {Han}}]{tout96}
{Tout}, C.~A., {Pols}, O.~R., {Eggleton}, P.~P., \& {Han}, Z. 1996, \mnras,
  281, 257

\bibitem[{{van Paradijs} \& {McClintock}(1995)}]{vm95}
{van Paradijs}, J. \& {McClintock}, J.~E. 1995, in X-Ray Binaries, 
  ed. W.~H.~G. {Lewin}, J.~{van Paradijs}, \& E.~P.~J. {van den Heuvel},
  58--125

\bibitem[{{Vrtilek} {et~al.}(1990){Vrtilek}, {Raymond}, {Garcia}, {Verbunt},
  {Hasinger}, \& {Kurster}}]{vrt+90}
{Vrtilek}, S.~D., {Raymond}, J.~C., {Garcia}, M.~R., {Verbunt}, F., {Hasinger},
  G., \& {Kurster}, M. 1990, \aap, 235, 162

\bibitem[{{Williams}(1980)}]{wil80}
{Williams}, R.~E. 1980, \apj, 235, 939

\bibitem[{{Williams} \& {Ferguson}(1982)}]{wf82}
{Williams}, R.~E. \& {Ferguson}, D.~H. 1982, \apj, 257, 672

\bibitem[{{Woudt} {et~al.}(2004){Woudt}, {Warner}, \& {Pretorius}}]{wou+04}
{Woudt}, P.~A., {Warner}, B., \& {Pretorius}, M.~L. 2004, \mnras, 351, 1015

\bibitem[{{York} {et~al.}(2000)}]{york+00}
{York}, D.~G. {et~al.} 2000, \aj, 120, 1579

\bibitem[{{Zacharias} {et~al.}(2004){Zacharias}, {Monet}, {Levine}, {Urban},
  {Gaume}, \& {Wycoff}}]{zac+04}
{Zacharias}, N., {Monet}, D.~G., {Levine}, S.~E., {Urban}, S.~E., {Gaume}, R.,
  \& {Wycoff}, G.~L. 2004, in Bulletin of the American Astronomical Society,
  Vol.~36, Bulletin of the American Astronomical Society, 1418--+

\end{thebibliography}

\begin{deluxetable}{l c c c c c c}
%%\tablecolumns{7}
\tablewidth{0pt}
\tablecaption{Hydrogen emission lines in J1023\label{tab:bal}}
\tablehead{
\colhead{Line}  & \colhead{$\lambda_{\rm vac}$} & \colhead{$\delta\lambda$} &
\colhead{$V$}  & \colhead{ FWHM}  & \colhead{ EW } & \colhead{ Flux/10$^{-16}$ } \\
	        & \colhead{ (\AA) } & \colhead{ (\AA) } &
\colhead{(km s$^{-1}$)} & \colhead{(km s$^{-1}$)} & \colhead{(\AA) }& 
\colhead{(ergs s$^{-1}$ cm$^{-2}$) } }
\startdata
H$\zeta$ $b$ & 3890.1 & $-8.9\pm$0.3 & $-690\pm23$ & 610$\pm$210 & 
3.01$\pm$0.22 & 78.0$\pm$5.7 \\
H$\zeta$ $r$ & 3890.1 & 6.3$\pm$0.4 & $488\pm31$  & 720$\pm$260 & 
2.76$\pm$0.23 & 71.1$\pm$5.8 \\
H$\epsilon$ $b$ & 3971.2 & $-9.4\pm$0.3 & $-709\pm22$ & 640$\pm$210  &
3.07$\pm$0.21 & 77.1$\pm$5.3 \\
H$\epsilon$ $r$ & 3971.2 & 6.3$\pm$0.3 & 479$\pm$24 & 660$\pm$220 &
2.96$\pm$0.21 & 74.0$\pm$5.2 \\
H$\delta$ $b$ & 4102.9 & $-9.1\pm$0.3 & $-669\pm22$ & 690$\pm$210 &
2.14$\pm$0.13 & 51.2$\pm$3.0 \\
H$\delta$ $r$ & 4102.9 & 10.0$\pm$0.3 & 728$\pm$25 & 1090$\pm$320 &
3.53$\pm$0.16 & 83.7$\pm$3.8 \\
H$\gamma$ $b$ & 4341.7 & $-9.3\pm0.2$ & $-639\pm16$ & 600$\pm$190 &
3.89$\pm$0.21 & 85.2$\pm$4.5 \\
H$\gamma$ $r$ & 4341.7 & 7.2$\pm$0.3 & 495$\pm$24 & 890$\pm$ 290 &
4.76$\pm$0.23 & 103.4$\pm$5.1 \\
H$\beta$ $b$ & 4862.7 & $-9.9\pm0.3$ & $-610\pm20$ & 780$\pm$250 &
5.09$\pm$0.25 & 93.0$\pm$4.5 \\
H$\beta$ $r$ & 4862.7 & 7.7$\pm$0.4 & 474$\pm$23 & 760$\pm$250 &
4.39$\pm$0.23 & 79.9$\pm$4.2 \\
H$\alpha$ $b$ & 6564.7 & $-9.5\pm0.4$ & $-432\pm19$ & 770$\pm$230 &
8.45$\pm$0.35 & 97.5$\pm$4.1 \\
H$\alpha$ $r$ & 6564.7 & $7.8\pm0.3$ & $355\pm15$ & 600$\pm$190 &
7.45$\pm0.29$ & 85.5$\pm$3.3 \\

\enddata
\end{deluxetable}

\begin{deluxetable}{l c c c c c c}
\tablecolumns{7}
\tablewidth{0pt}
\tablecaption{Helium emission lines in J1023\label{tab:he}}
\tablehead{
\colhead{Line}  & \colhead{$\lambda_{\rm vac}$} & \colhead{$\delta\lambda$} &
\colhead{$V$}  & \colhead{ FWHM}  & \colhead{ EW } & \colhead{ Flux/10$^{-16}$ } \\
	        & \colhead{ (\AA) } & \colhead{ (\AA) } &
\colhead{(km s$^{-1}$)} & \colhead{(km s$^{-1}$)} & \colhead{(\AA) }& 
\colhead{(ergs s$^{-1}$ cm$^{-2}$) } }
\startdata
\ion{He}{1} $b$ & 4027.3 & $-6.3\pm$0.5 & $-472\pm34$ & 910$\pm$300 & 
3.22$\pm$0.22 & 79.2$\pm$5.5 \\
\ion{He}{1} $r$ & 4027.3 & 10.1$\pm$0.8 & $755\pm58$  & 580$\pm$300 & 
0.96$\pm$0.18 & 23.4$\pm$4.4 \\
\ion{He}{1} $b$ & 4472.7 & $-8.1\pm$0.7 & $-541\pm48$ & 580$\pm$290  &
1.33$\pm$0.29 & 27.8$\pm$6.0 \\
\ion{He}{1} $r$ & 4472.7 & 4.9$\pm$1.2 & 327$\pm$83 & 1500$\pm$530 &
5.22$\pm$0.33 & 108.2$\pm$6.9 \\
\ion{He}{2} $b$ & 4687.0 & $-10.2\pm$0.3 & $-650\pm21$ & 430$\pm$160 &
1.69$\pm$0.17 & 32.7$\pm$3.3 \\
\ion{He}{2} $r$ & 4687.0 & 9.2$\pm$0.8 & 591$\pm$49 & 1340$\pm$470 &
3.99$\pm$0.26 & 76.8$\pm$5.0 \\
\ion{He}{1} $b$ & 4923.3 & $-9.1\pm0.6$ & $-550\pm33$ & 640$\pm$240 &
1.94$\pm$0.20 & 34.9$\pm$3.5 \\
\ion{He}{1} $r$ & 4923.3 & 8.5$\pm$0.5 & 519$\pm$29 & 520$\pm$200 &
1.74$\pm$0.18 & 31.0$\pm$3.2 \\
\ion{He}{1} $b$ & 5017.1 & $-7.6\pm0.6$ & $-452\pm35$ & 700$\pm$260 &
2.29$\pm$0.22 & 39.8$\pm$3.8 \\
\ion{He}{1} $r$ & 5017.1 & 11.5$\pm$0.5 & 691$\pm$29 & 870$\pm$290 &
3.98$\pm$0.23 & 68.9$\pm$4.0 \\
\ion{He}{1} $b$ & 5877.3 & $-7.6\pm0.6$ & $-388\pm31$ & 720$\pm$240 &
3.82$\pm$0.26 & 52.4$\pm$3.6 \\
\ion{He}{1} $r$ & 5877.3 & $5.9\pm0.3$ & $300\pm17$ & 350$\pm$130 &
2.40$\pm0.19$ & 32.8$\pm$2.6 \\
\ion{He}{1} $b$ & 6680.0 & $-10.7\pm1.0$ & $-482\pm45$ & 800$\pm$300 &
2.80$\pm$0.27 & 31.2$\pm$3.0 \\
\ion{He}{1} $r$ & 6680.0 & $9.5\pm0.9$ & $427\pm40$ & 630$\pm$260 &
2.25$\pm0.23$ & 25.0$\pm$2.6 \\
\ion{He}{1} $b$ & 7067.1 & $-10.0\pm1.0$ & $-425\pm44$ & 620$\pm$250 &
1.98$\pm$0.25 & 20.1$\pm$2.6 \\
\ion{He}{1} $r$ & 7067.1 & $9.1\pm0.9$ & $388\pm39$ & 560$\pm$240 &
1.99$\pm0.23$ & 20.1$\pm$2.4 \\

%%\tablenotetext{a}{Days since MJD 54597.0.}
%%\tablenotetext{b}{1$\sigma$ uncertainty resulting from PSF fitting.}
\enddata
\end{deluxetable}

\end{document}